\documentclass[aps, amsmath, amssymb, aps, pre, twocolumn,groupedaddress,superscriptaddress,nofootinbib,longbibliography]{revtex4-1}

\usepackage{multirow}
\usepackage{enumitem}
\setlist{noitemsep,topsep=0pt,parsep=0pt,partopsep=0pt}
\usepackage{array}
\usepackage[english]{babel}
\usepackage[colorlinks,linkcolor=black,citecolor=blue,urlcolor=black]{hyperref}
\usepackage{amsmath}
\usepackage{tabularx}

\usepackage{graphics}
\usepackage{graphicx}
\usepackage{xcolor}
\usepackage[config, labelfont={sf,bf}, textfont=sf]{subfig}
\graphicspath{{}}
\usepackage[suffix=, outdir=./figs/]{epstopdf}
\usepackage{hhline}

\usepackage{lineno}

\modulolinenumbers[5]

\DeclareMathOperator{\LEs}{LE}

\begin{document}

\title{A lower-bound estimate of the Lyapunov dimension \\
for the global attractor of the Lorenz system}

\author{N. V. Kuznetsov}
\email[]{Corresponding author: nkuznetsov239@gmail.com}
\affiliation{Faculty of Mathematics and Mechanics, St. Petersburg State University,
Peterhof, St. Petersburg, Russia}
\affiliation{Faculty of Information Technology,
University of Jyv\"{a}skyl\"{a}, Jyv\"{a}skyl\"{a}, Finland}
\affiliation{Institute for Problems in Mechanical Engineering RAS, Russia}
\author{T. N. Mokaev}
\affiliation{Faculty of Mathematics and Mechanics, St. Petersburg State University,
Peterhof, St. Petersburg, Russia}
\author{R. N. Mokaev}
\affiliation{Faculty of Mathematics and Mechanics, St. Petersburg State University,
Peterhof, St. Petersburg, Russia}
\affiliation{Faculty of Information Technology,
University of Jyv\"{a}skyl\"{a}, Jyv\"{a}skyl\"{a}, Finland}
\author{O. A. Kuznetsova}
\affiliation{Faculty of Mathematics and Mechanics, St. Petersburg State University,
Peterhof, St. Petersburg, Russia}
\author{E. V. Kudryashova}
\affiliation{Faculty of Mathematics and Mechanics, St. Petersburg State University,
Peterhof, St. Petersburg, Russia}

\date{\today}


\begin{abstract}
In this short report, for the classical Lorenz attractor
we demonstrate the applications of
the Pyragas time-delayed feedback control technique
and Leonov analytical method for
the Lyapunov dimension estimation
and verification of the Eden's conjecture.
The problem of reliable numerical computation of the
finite-time Lyapunov dimension
along the trajectories over large time intervals is discussed.
\end{abstract}

\maketitle

\section{Lorenz attractor and Pyragas stabilization of embedded unstable periodic orbits}

Consider the classical Lorenz system~\cite{Lorenz-1963}
\begin{equation}\label{eq:lorenz}
    \begin{cases}
        \dot{x} = - \sigma (x - y), \\
        \dot{y} = r x - y - x z, \\
        \dot{z} = - b z + xy,
    \end{cases}
\end{equation}
with physically sound parameters $\sigma, r > 0$, and $b \in [0,4]$.
For $r < 1$ it has only one globally stable equilibrium $S_0 = \big(0, \, 0, \, 0\big)$,
and for $r > 1$ the equilibrium $S_0$ turns into a saddle, while two new symmetric equilibria appear:
\begin{equation}\label{eq:lorenz:equilibria}
S_\pm = \big(\pm \sqrt{b (r-1)}, \, \pm \sqrt{b (r-1)}, \, r-1\big),
\end{equation}
which stability depends on the values of parameters.

System \eqref{eq:lorenz} is dissipative in the sense of Levinson (see e.g.~\citep{LeonovKM-2015-EPJST}),
i.e. there exist a global bounded absorbing set containing global attractor
$\mathcal{A}_{\rm glob}$, and
in some cases this attractor exhibits chaotic behavior.
For some values of parameters, it is possible to observe a case of multistability, when
the global attractor consists of several local attractors.
To get a visualization of such attractors
one needs to choose an initial point in the basin of attraction of a particular attractor
and observe how the trajectory, starting from this initial point,
after a transient process visualizes the attractor:
an attractor is called a \emph{self-excited attractor}
if its basin of attraction
intersects with any open neighborhood of an equilibrium,
otherwise, it is called a \emph{hidden attractor}
\citep{LeonovKV-2011-PLA,LeonovK-2013-IJBC,LeonovKM-2015-EPJST,Kuznetsov-2016}.
It was discovered numerically by E.~Lorenz that
in the phase space of system \eqref{eq:lorenz}
with parameters $r = 28$, $\sigma = 10$, $b = 8/3$
there exist a \emph{chaotic attractor} $\mathcal{A}$,
which is self-excited with respect to all equilibria $S_0$, $S_\pm$.

\begin{figure*}[ht]
  \centering
  \subfloat[]{
    \label{fig:lorenz:upo:all}
  \includegraphics[width=0.45\textwidth]{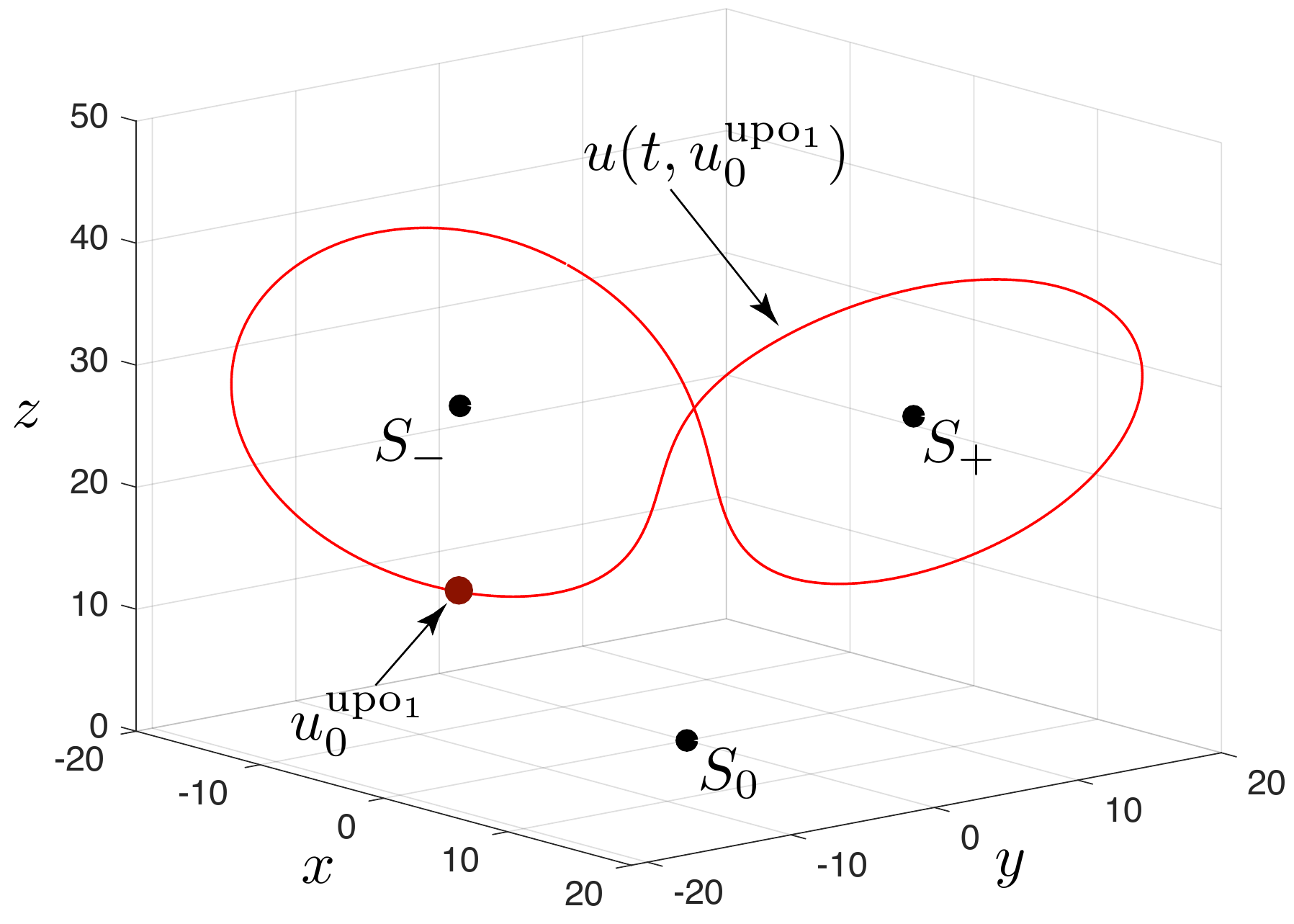}
  }\quad
  \subfloat[]{
    \label{fig:lorenz:upo:upo1attr}
    \includegraphics[width=0.45\textwidth]{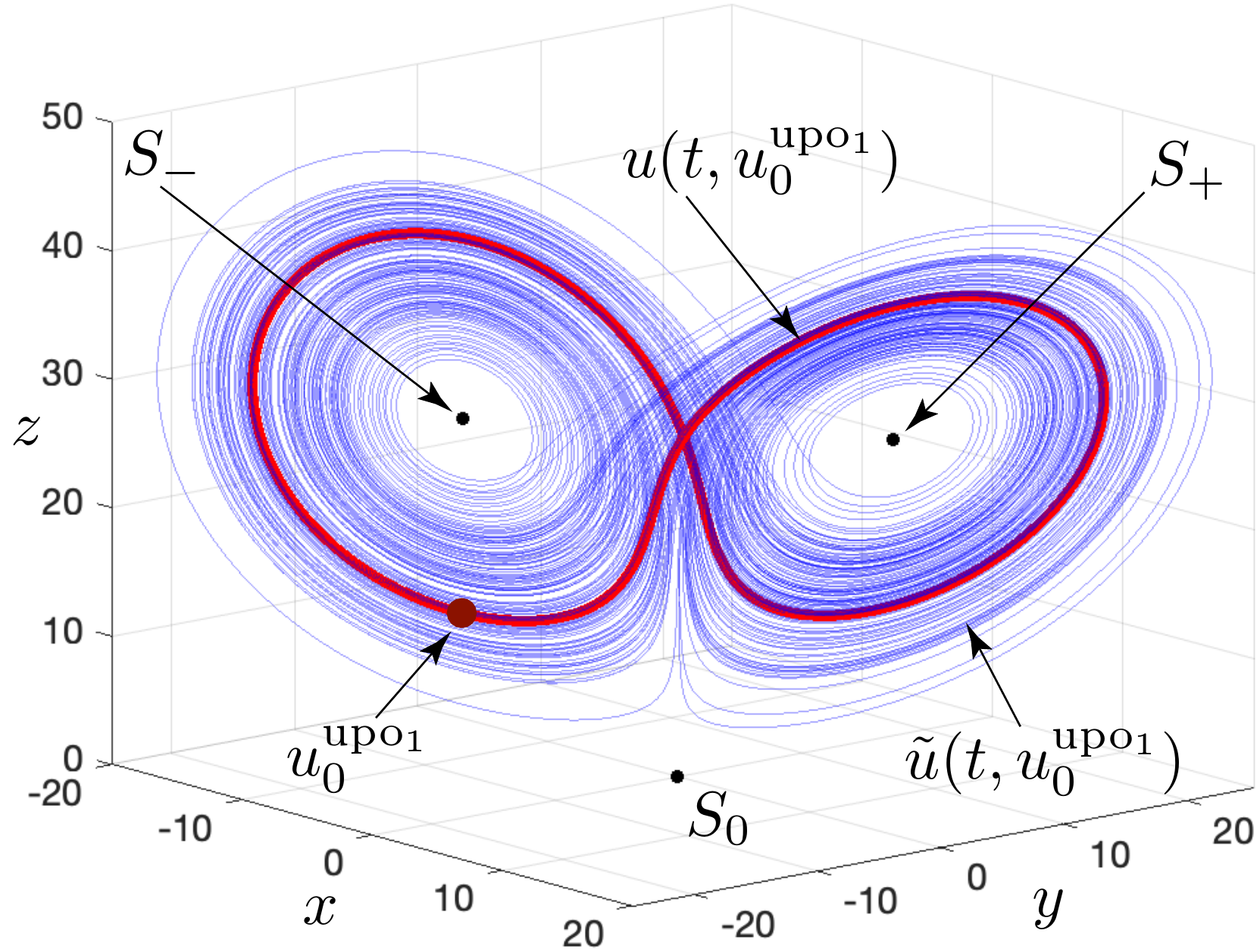}
  }
  \caption{Period-1 UPO $u^{\rm upo_1}(t)$ (red, period $\tau_1 = 1.5586$)
  stabilized using UDFC method,
  and pseudo-trajectory $\tilde{u}(t, u^{\rm upo_1}_0)$ (blue, $t \in [0,100]$)
  in system \eqref{eq:lorenz} with parameters $r = 28$, $\sigma = 10$, $b = 8/3$.
  }
  \label{fig:lorenz:upo}
\end{figure*}

The ''skeleton'' of a chaotic attractor comprises
embedded unstable periodic orbits (UPOs)
(see e.g. \citep{AfraimovicBSh-1977,AuerbachCEGP-1987,Cvitanovic-1991}), and
one of the effective methods among others for the computation of UPOs
is the \emph{delay feedback control} (DFC) approach,
suggested by K.~Pyragas \citep{Pyragas-1992}
(see also discussions in \citep{KuznetsovLS-2015-IFAC,ChenY-1999,LehnertHFGFS-2011}).
This approach allows Pyragas and his progeny
to stabilize and study UPOs in various chaotic dynamical systems.
Nevertheless, some general analytical results have been obtained~\cite{HootonA-2012},
showing that DFC has a certain limitation, called the odd number limitation (ONL),
which is connected with an odd number of real Floquet multipliers larger
than unity.
In order to overcome ONL, later Pyragas suggested a modification of the classical DFC
technique, which was called the unstable delayed feedback control (UDFC)~\cite{Pyragas-2001}.

Rewrite system \eqref{eq:lorenz} in a general form
\begin{equation}\label{eq:sys_gen}
  \dot{u} = f(u).
\end{equation}
Let $u^{\rm upo}(t,u^{\rm upo_1}_0)$ be its UPO with period $\tau > 0$,
$u^{\rm upo}(t - \tau,u^{\rm upo_1}_0) = u^{\rm upo}(t,u^{\rm upo_1}_0)$,
and initial condition $u^{\rm upo_1}_0=u^{\rm upo}(0,u^{\rm upo_1}_0)$.
To compute the UPO and overcome ONL, we add the UDFC in the following form:
\begin{equation}\label{eq:closed_loop_syst}
\begin{aligned}
\dot{u}(t) &= f(u(t)) + K B \, \big[F_N(t) + w(t)\big], \\
\dot{w}(t) &= \lambda_c^0 w(t) + (\lambda_c^0 - \lambda_c^\infty) F_N(t), \\
F_N(t) &= C^*u(t) - (1\!-\!R) \sum_{k=1}^N R^{k-1} C^*u(t - k T),
\end{aligned}
\end{equation}
where
$0 \leq R < 1$ is an extended DFC parameter,
$N = 1,2,\ldots,\infty$ defines the number of previous states involved in
delayed feedback function $F_N(t)$,
$\lambda_c^0 > 0$, and $\lambda_c^\infty < 0$ are
additional unstable degree of freedom parameters,
$B, C$ are vectors and $K > 0$ is a feedback gain.
For initial condition $u^{\rm upo_1}_0$ and $T = \tau$ we have
\[
  F_N(t) \equiv 0, \ w(t) \equiv 0,
\]
and, thus, the solution of system \eqref{eq:closed_loop_syst}
coincides with the periodic solution
of initial system \eqref{eq:sys_gen}.

For the Lorenz system~\eqref{eq:lorenz} with parameters $r = 28$, $\sigma = 10$, $b = 8/3$
using~\eqref{eq:closed_loop_syst} with
$B^* = \left(0, 1, 0\right)$, $C^* = \left(0, 1, 0\right)$,
$R = 0.7$, $N =100$, $K = 3.5$, $\lambda^0_c = 0.1$, $\lambda^\infty_c = -2$,
one can stabilize a period-1 UPO $u^{\rm upo_1}(t,u_0)$
with period $\tau_1 = 1.5586$
from the initial point $u_0 = (1, 1, 1)$, $w_0 = 0$
(see Fig.~\ref{fig:lorenz:upo}).
Results of this experiment could be repeated using various other numerical approaches
(see e.g. \cite{Viswanath-2001,Budanov-2018,PchelintsevPY-2019-LorenzUpo}),
and are in agreement with similar results on the existence of
UPOs embedded in the Lorenz attractor \cite{GaliasT-2008-LorenzUpo,BarrioDT-2015-LorenzUpo}.
However, the Pyragas procedure, in general, is more convenient for UPOs numerical visualization.

For the initial point $u^{\rm upo_1}_0 \approx (-6.2262, -11.0027, 13.0515)$
on the UPO $u^{\rm upo_{1}}(t) = u(t, u^{\rm upo_1}_0)$
we numerically compute the trajectory
of system \eqref{eq:closed_loop_syst} without the stabilization (i.e. with $K = 0$)
on the time interval $[0,T=100]$ (see Fig.~\ref{fig:lorenz:upo:upo1attr}).
We denote it by $\tilde{u}(t, u^{\rm upo_1}_0)$
to distinguish this pseudo-trajectory from the periodic orbit $u(t, u^{\rm upo_1}_0)$.
One can see that on the initial small time interval $[0,T_1 \approx 11]$,
even without the control,
the obtained trajectory $\tilde{u}(t, u^{\rm upo_1}_0)$
traces approximately the ''true'' periodic orbit $u(t, u^{\rm upo_1}_0)$.
But for $t > T_1$, without a control,
the trajectory $\tilde{u}(t, u^{\rm upo_1}_0)$
diverge from $u^{\rm upo_1}(t, u^{\rm upo_1}_0)$ and visualize a local chaotic attractor~$\mathcal{A}$.

Remark that in numerical computation of trajectory over a finite-time interval
it is also difficult to distinguish a \emph{sustained chaos} from
a \emph{transient chaos}
(a transient chaotic set in the phase space,
which can persist for a long time) \cite{GrebogiOY-1983}.
This challenging task is related to an open problem about the existence of
a hidden chaotic attractor in the Lorenz system~\eqref{eq:lorenz}
(see e.g. discussions in
\cite{LeonovK-2015-AMC,LeonovKM-2015-EPJST,ChenKLM-2017-IJBC,SprottM-2018-HA}).

\section{Lyapunov dimension estimation and Eden conjecture}


Following \cite{Kuznetsov-2016-PLA,KuznetsovLMPS-2018}, let us outline
the concept of the \emph{finite-time Lyapunov dimension},
which is convenient for carrying out numerical experiments with finite time.

For a fixed $t\geq 0$ let us consider the map $u(t,\cdot): \mathbb{R}^3 \to \mathbb{R}^3$
defined by the shift operator along the solutions of system~\eqref{eq:lorenz}:
$u(t,u_0)$, $u_0 \in \mathbb{R}^3$.
Since system \eqref{eq:lorenz} possesses an absorbing set,
the existence and uniqueness of solutions of system~\eqref{eq:lorenz}
for $t \in [0,+\infty)$ take place and, therefore, the system generates a \emph{dynamical system}
$\big(\{u(t,\cdot)\}_{t\geq0}, (\mathbb{R}^3, |\cdot|)\big)$.

Consider linearization of system~\eqref{eq:lorenz}
along the solution $u(t,u_0)$
and its $3\!\times\!3$ fundamental matrix of solutions $\Phi(t,u_0)$:
$\dot \Phi(t,u_0) = Df((u(t,u_0)) \Phi(t,u_0)$,
where
$\Phi(0,u_0)~=~I$ is a unit $3\!\times\!3$ matrix.
Denote by $\sigma_i(t,u_0) = \sigma_i(\Phi(t,u_0))$, $i = 1,2,3$,
the singular values of $\Phi(t,u_0)$
(i.e. the square roots of
the eigenvalues of the symmetric matrix $\Phi(t,u_0)^*\Phi(t,u_0)$
with respect \noindent to their algebraic multiplicity)\footnote{
Symbol $^*$ denotes the transposition of matrix.
},
ordered so that $\sigma_1(t,u_0) \geq \sigma_2(t,u_0) \geq \sigma_3(t,u_0) > 0$
for any $u_0 \in \mathbb{R}^3$ and $t > 0$.

Consider a set of \emph{finite-time Lyapunov exponents}
at the point $u_0$:
\begin{equation}\label{ftLE}
  \LEs_i(t,u_0) = \frac{1}{t}\ln\sigma_i(t,u_0), \ t > 0, \quad i=1,2,3.
\end{equation}
Here, the set $\{\LEs_i(t,u_0)\}_{i=1}^3$ is ordered by decreasing
(i.e. $\LEs_1(t,u_0) \geq \LEs_2(t,u_0) \geq \LEs_3(t,u_0)$ for all $t>0$).
The \emph{finite-time local Lyapunov dimension} \cite{Kuznetsov-2016-PLA,KuznetsovLMPS-2018}
can be defined via an analog of the \emph{Kaplan-Yorke formula}
with respect to the set of ordered finite-time Lyapunov exponents $\{\LEs_i(t,u_0)\}_{i=1}^3$:
\begin{multline}\label{lftKY}
   \dim_{\rm L}(t, u_0)\!=\!
    j(t,u_0) + \tfrac{\LEs_1\!(t,u_0) + \cdot\cdot + \LEs_{j(,u_0)}\!(t,u_0)
   }{|\LEs_{j(t,u_0)\!+\!1}(t,u_0)|},
\end{multline}
where $j(t,u_0) = \max\{m: \sum_{i=1}^{m}\LEs_i(t,u_0) \geq 0\}$.
Then the \emph{finite-time Lyapunov dimension}
of dynamical system with respect to a set $\mathcal{A}$ is defined as:
\begin{equation}\label{DOmaptmax}
  \dim_{\rm L}(t, \mathcal{A}) = \sup\limits_{u_0 \in \mathcal{A}} \dim_{\rm L}(t, u_0).
\end{equation}

The \emph{Douady--Oesterl\'{e} theorem} \cite{DouadyO-1980} implies that
for any fixed $t > 0$
the finite-time Lyapunov dimension on a compact invariant set $\mathcal{A}$,
defined by \eqref{DOmaptmax},
is an upper estimate of the Hausdorff dimension:
$\dim_{\rm H} \mathcal{A} \leq \dim_{\rm L}(t, \mathcal{A})$.
The best estimation is called the \emph{Lyapunov dimension} \cite{Kuznetsov-2016-PLA}
\begin{multline*}
   \dim_{\rm L} \mathcal{A}
   = \inf_{t > 0}\sup\limits_{u_0 \in \mathcal{A}} \dim_{\rm L}(t, u_0) = \\
   = \liminf_{t \to +\infty}\sup\limits_{u_0 \in \mathcal{A}} \dim_{\rm L}(t, u_0).
\end{multline*}

We use the {\it adaptive algorithm for the computation of
the finite-time Lyapunov dimension and exponents} for trajectories on
the local attractor $\mathcal{A}$ \cite{KuznetsovLMPS-2018}.
In order to distinguish the corresponding values for
the stabilized UPO $u(t,u^{\rm upo_1}_0)$ with a period $\tau_1 = 1.5586$
and for the pseudo-trajectory $\tilde u(t, u^{\rm upo_1}_0)$
computed without Pyragas stabilization in
our experiment we use the following notations
for finite-time Lyapunov dimensions:
$\dim_{\rm L}(u(t, \cdot), u^{\rm upo_1}_0)$ and
$\dim_{\rm L}(\tilde u(t, \cdot), u^{\rm upo_1}_0)$,
respectively.
\begin{figure*}[ht!]
 \centering
 \subfloat[]{
    \label{fig:lorenz:upo1:attr}
    \includegraphics[width=0.45\textwidth]{lorenz_UPO-LR_ch}
  }\quad
 \subfloat[]{
    \label{fig:lorenz:upo1:LD}
    \includegraphics[width=0.45\textwidth]{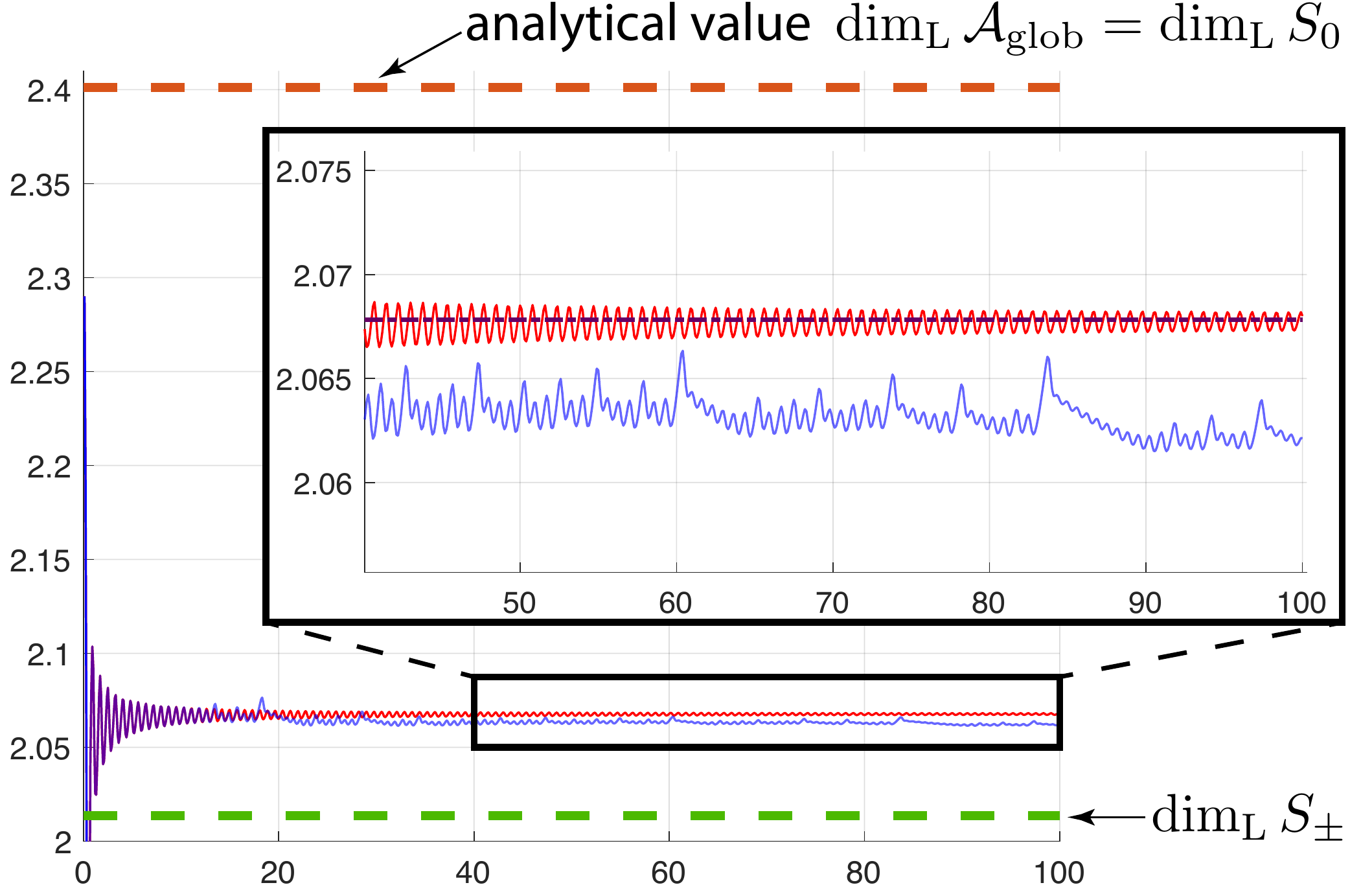}
  }
\caption{Evolution of FTLDs $\dim_{\rm L}(u(t, \cdot), u^{\rm upo_1}_0)$ (red) and
$\dim_{\rm L}(\tilde u(t, \cdot), u^{\rm upo_1}_0)$ (blue) computed
on the time interval $t \in [0,100]$
along the UPO $u^{\rm upo_1}(t) = u (t, u^{\rm upo_1}_0)$ (red)
and the trajectory $\tilde u (t, u^{\rm upo_1}_0)$ (blue) integrated without stabilization, respectively.
Both trajectories start from the point $u^{\rm upo_1}_0 = (-6.2262, -11.0027, 13.0515)$.
}
\label{fig:lorenz:upoLD}
\end{figure*}
\begin{figure*}[!ht]
  \centering
  \includegraphics[width=0.9\textwidth]{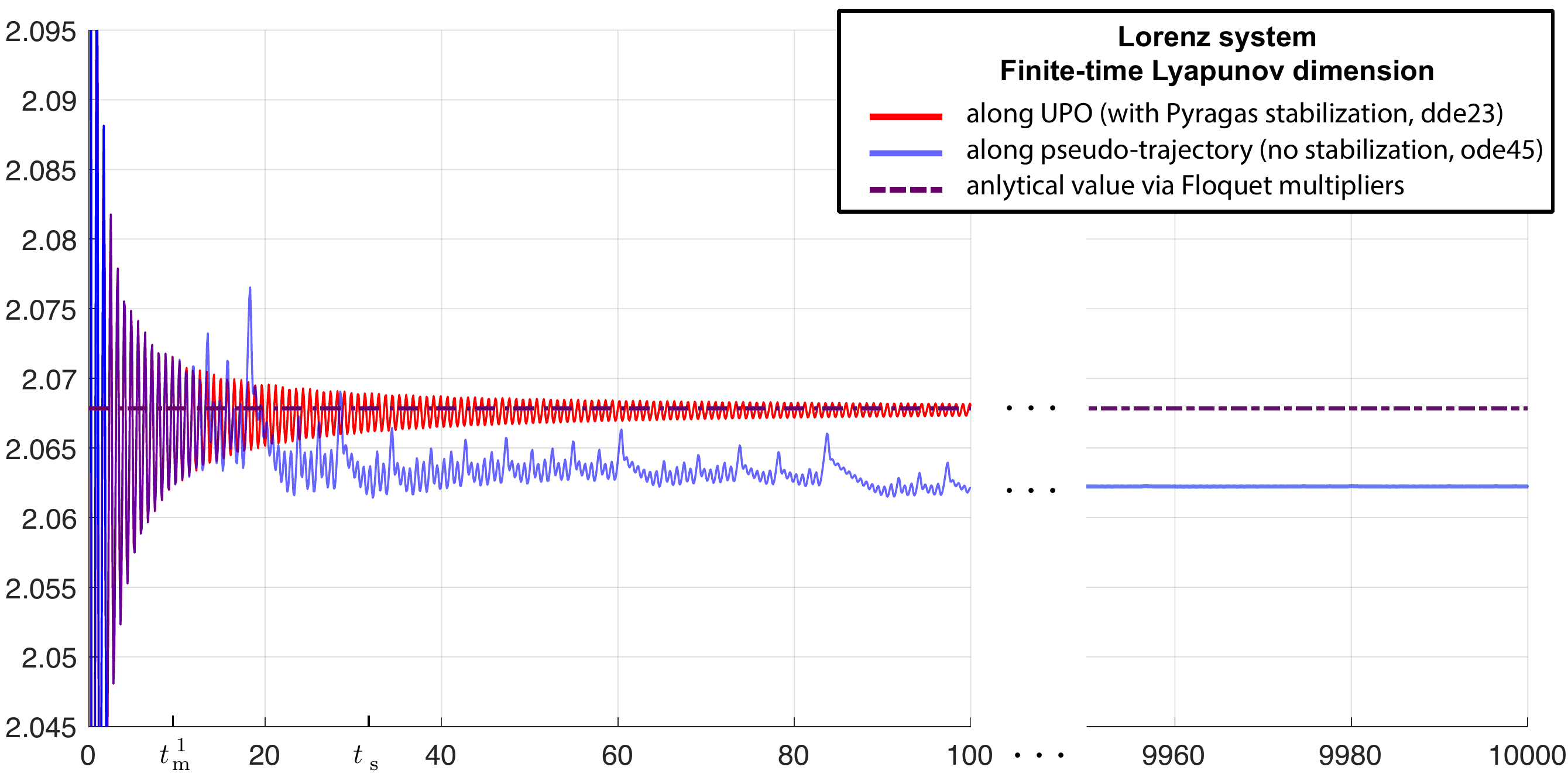}
  \caption{Evolution of FTLDs $\dim_{\rm L}(u(t, \cdot), u^{\rm upo_1}_0)$ (red) and
    $\dim_{\rm L}(\tilde u(t, \cdot), u^{\rm upo_1}_0)$ (blue) computed
    on the long time interval $t \in [0,10000]$
    along the UPO $u^{\rm upo_1}(t) = u (t, u^{\rm upo_1}_0)$ (red)
    and the trajectory $\tilde u (t, u^{\rm upo_1}_0)$ (blue) integrated without stabilization, respectively.
    Both trajectories start from the point $u^{\rm upo_1}_0 = (-6.2262, -11.0027, 13.0515)$.}
  \label{fig:lorenz:upoLD:long}
\end{figure*}

The comparison of the obtained values of finite-time Lyapunov dimensions
computed along the stabilized UPO and the trajectory without stabilization
gives us the following results.
On the initial small part of the time interval, one can indicate the coincidence of these values
with a sufficiently high accuracy.
For the UPO and for the unstabilized trajectory
the finite-time local Lyapunov dimensions
$\dim_{\rm L}(u(t, \cdot), u^{\rm upo_1}_0)$ and
$\dim_{\rm L}(\tilde u(t, \cdot), u^{\rm upo_1}_0)$
coincide up to the 4th decimal place inclusive on the interval $[0,t_m^1 \approx 7\tau_1]$.
After $t > t_m^1$ the difference in values becomes significant and the
corresponding graphics diverge
in such a way that the part of the graph corresponding to the unstabilized trajectory
is lower than the part of the graph corresponding to the UPO
(see Fig.~\ref{fig:lorenz:upo1:LD}, Fig.~\ref{fig:lorenz:upoLD:long}).

The Jacobi matrix at the saddle-foci equilibria $S_\pm$ has simple eigenvalues,
which give the following: $\dim_{\rm L} S_\pm  = 2.0136$.
The UPO $u^{\rm upo_1}$ with period $\tau_1 = 1.5586$
has the following Floquet multipliers:
$\rho_1 = 4.7127$, $\rho_2~=~1$, $\rho_3 = -1.19 \cdot 10^{-10}$ and
corresponding Lyapunov exponents: $\{ \frac{1}{\tau_1} \log \rho_i \}_{i=1}^3$.
Thus, for the local Lyapunov dimension of this UPO
we obtain:
$\dim_{\rm L} u^{\rm upo_1} = 2.0678 \lessapprox 2.0679 = \dim_{\rm L}(u(100, \cdot), u^{\rm upo_1}_0)$.

Using an effective analytical technique,
proposed by Leonov~\citep{Leonov-1991-Vest,Kuznetsov-2016-PLA},
which is based on a combination of the Douady-Oesterl\'{e} approach and
the direct Lyapunov method,
it is possible to obtain \cite{LeonovKKK-2016-CNSCS,Leonov-2018-UMZh}
the exact formula of the Lyapunov dimension for
the global attractor~$\mathcal{A}_{\rm glob}$
of the Lorenz system \eqref{eq:lorenz}:
\begin{equation}\label{thm:LD:estim}
  \dim_{\rm L} \mathcal{A}_{\rm glob}
  = 3 - \tfrac{2(\sigma+b+1)}{\sigma + 1 + \sqrt{(\sigma-1)^2 + 4 \sigma r}}
\end{equation}
for the case, when $r \sigma > (\sigma + b) (b + 1)$.

\section{Conclusion}
In this note, for the Lorenz system \eqref{eq:lorenz}
with classical values of parameters
$r = 28$, $\sigma = 10$, $b = 8/3$
we have studied the Eden conjecture~\cite[p.98]{Eden-1989-PhD}
and obtained the following relations:
\begin{equation*}\label{}
\begin{aligned}
  \dim_{\rm L} \mathcal{A_{\rm glob}} = \dim_{\rm L} S_0 =
    3 - \tfrac{2(\sigma+b+1)}{\sigma + 1 + \sqrt{(\sigma-1)^2 + 4 \sigma r}}\!=\!2.4013 > \\ >
    \dim_{\rm L} \mathcal{A} \geq
    \dim_{\rm L} {u}^{\rm upo_1} = 2.0678 >
    \dim_{\rm L}(\tilde u(100, \cdot), u^{\rm upo_1}_0) \\
    = 2.0621 > \dim_{\rm L} S_\pm = 2.0136.
\end{aligned}
\end{equation*}

Here, since the global Lorenz attractor contains a period-1 UPO:
$\mathcal{A}_{\rm glob} \supset u^{\rm upo_1}$,
we have the following lower-bound estimate
for the Lyapunov dimension:
\(\dim_{\rm L} \mathcal{A}_{\rm glob} \geq 2.0678 = \dim_{\rm L} {u}^{\rm upo_1}\).
Similar experiment and results for the R\"{o}ssler system~\citep{Rossler-1976}
are presented in \cite{KuznetsovM-2019-AIP,KuznetsovKKMD-2019-IFAC}.

Concerning the time of integration, remark that while the time series obtained from a \emph{physical experiment} are assumed to be reliable on the whole considered time interval,
the time series produced by the integration of
\emph{mathematical dynamical model}
can be reliable on a limited time interval only
due to computational errors
(caused by finite precision arithmetic and numerical integration of ODE).
Thus, in general, the closeness of the real trajectory $u(t,u_0)$
and the corresponding pseudo-trajectory $\tilde u(t,u_0)$
calculated numerically can be guaranteed on a limited short time interval only.

In our experiment, if we continue computation over a long time interval $[0,10000]$
of FTLD along
the stabilized UPO and the pseudo-trajectory obtained without Pyragas stabilization,
as a result, completely different values will be obtained (see Fig.~\ref{fig:lorenz:upoLD:long}).
Evolution of $\dim_{\rm L}(u(t, \cdot), u^{\rm upo_1}_0)$ along the stabilized UPO will tend
to the analytical value $\dim_{\rm L} {u}^{\rm upo_1} = 2.0678$, computed via Floquet multipliers,
while evolution of $\dim_{\rm L}(\tilde u(t, \cdot), u^{\rm upo_1}_0)$ along the pseudo-trajectory will
converge to the value $2.0622$\footnote{
    The following results on the dimension of the Lorenz attractor
    with parameters $r=28$,$\sigma=10$,$b=8/3$ can be found in the literature.
    In
    \citep[p.~193]{GrassbergerP-1983} and \citep[p.~3529]{BenziPPV-1984}
    the fractal (box-counting, capacity) dimension
    is estimated as $\mathbf{2.06 \pm 0.01}$.
    For the correlation dimension the following results are known:
    $\mathbf{2.05 \pm 0.01}$ in \citep[p.~193]{GrassbergerP-1983} and \citep[p.~456]{Strogatz-1994};
    $\mathbf{2.06 \pm 0.03}$ in \citep[p.~47]{MalinetskiiP-1988};
    $\mathbf{2.049 \pm 0.096}$ in \citep[p.~1874]{SprottR-2001};
    $\mathbf{2.05}$ in \citep[p.~80]{Fuchs-2013}.
    For the Lyapunov dimension the following values have been computed:
    $\mathbf{2.063}$ in \citep[p.~92]{Lorenz-1984} and \citep[p.~1957]{NeseDW-1987};
    $\mathbf{2.05}$ in \citep[p.~267]{DoeringG-1995};
    $\mathbf{2.062}$ in \citep[p.~1874]{SprottR-2001}, \citep[p.~115]{Sprott-2003} and \citep[p.~53]{Lappa-2009};
    $\mathbf{2.06215}$ \citep[p.~033124-3]{Sprott-2007} and \citep[p.~83]{Fuchs-2013}.
    Also, let us mention estimates for the global attractor:
    $2.401 \leq \dim_{\rm L} \mathcal{A_{\rm glob}} \leq 2.409$ \citep[p.~170]{EdenFT-1991} and
    $\dim_{\rm L} \mathcal{A_{\rm glob}} \approx 2.401...$ in \citep[p.~267]{DoeringG-1995}.
}.
These results are in good agreement with the
rigorous analysis of the time interval choices for reliable numerical computation of trajectories
for the Lorenz system:
the time interval for reliable computation
with 16~significant digits and error $10^{-4}$
is estimated as $[0, 36]$, with error $10^{-8}$
is estimated as $[0, 26]$ (see \cite{KehletL-2013,KehletL-2017}),
and reliable computation for a longer time interval, e.g. $[0,10000]$ in \cite{LiaoW-2014},
is a challenging task that requires significant increase of the precision of the floating-point
representation and the use of supercomputers.
Analytical aspects of this problem are related to
the shadowing theory (see e.g. \cite{Pilyugin-2011}).

\section*{Acknowledgement}
This work was supported by the Russian Science Foundation 19-41-02002.


\end{document}